\documentclass[twocolumn,showpacs,preprintnumbers,amsmath,amssymb,prb]{revtex4}

\usepackage{graphicx,psfrag}
\usepackage{dcolumn}
\usepackage{bm,bbm}
\usepackage{pstricks,pst-plot}
\def\d{{\mathrm d}}

\newcommand{\ppfq}[2]{\frac{\partial^{2} #1}{\partial #2^{2}}}

\newcommand{\mbf}[1]{{\mathbf #1}}

\begin{document}


\title{Microscopic non-equilibrium theory of quantum well solar cells}

\author{U. Aeberhard}
\email{urs.aeberhard@psi.ch}
\author{R. H. Morf}
\affiliation{Condensed Matter Theory, Paul Scherrer Institute,CH-5232 Villigen,
Switzerland}

\date{\today}

\begin{abstract}
  We present a microscopic theory of bipolar quantum well structures in the
  photovoltaic regime,
  based on the non-equilibrium Green's function formalism for a multi band
  tight binding Hamiltonian. The quantum kinetic equations for the single
  particle Green's functions of electrons and holes are
  self-consistently coupled to Poisson's equation, including inter-carrier
  scattering on the Hartree level. Relaxation and broadening mechanisms 
  are considered by the inclusion of acoustic and optical electron-phonon interaction in 
  a self consistent Born approximation of the scattering self energies.
  Photogeneration of carriers is described on the same level in terms of a self
  energy derived from the standard dipole approximation of
  the electron-photon interaction. Results from a simple two band
  model are shown for the local
  density of states, spectral response, current spectrum, and
  current-voltage characteristics for generic single quantum well systems.
  
\end{abstract}

\maketitle

\section{\label{sec:level1}Introduction} 
\begin{figure}[h!]
\begin{minipage}{5.8cm}
	\psfrag{p}[][]{p}
	\psfrag{i}[][]{i}
	\psfrag{n}[][]{n}
	\psfrag{electrons}[][r]{electrons}
	\psfrag{holes}[][r]{holes}
	\psfrag{z}[][r]{z}
	\psfrag{E}[][r]{E}
	\psfrag{V}[][r]{\quad V$_{bias}$}
	\psfrag{ML}[][]{$\mu_{L}$}
	\psfrag{MR}[][]{$\mu_{R}$}
	\psfrag{1}{1}
	\psfrag{2}{2}
	\psfrag{3}{3}
	\psfrag{4}{4}
	\psfrag{5}{5}
	\psfrag{6}{6}
     \includegraphics[width=7cm]{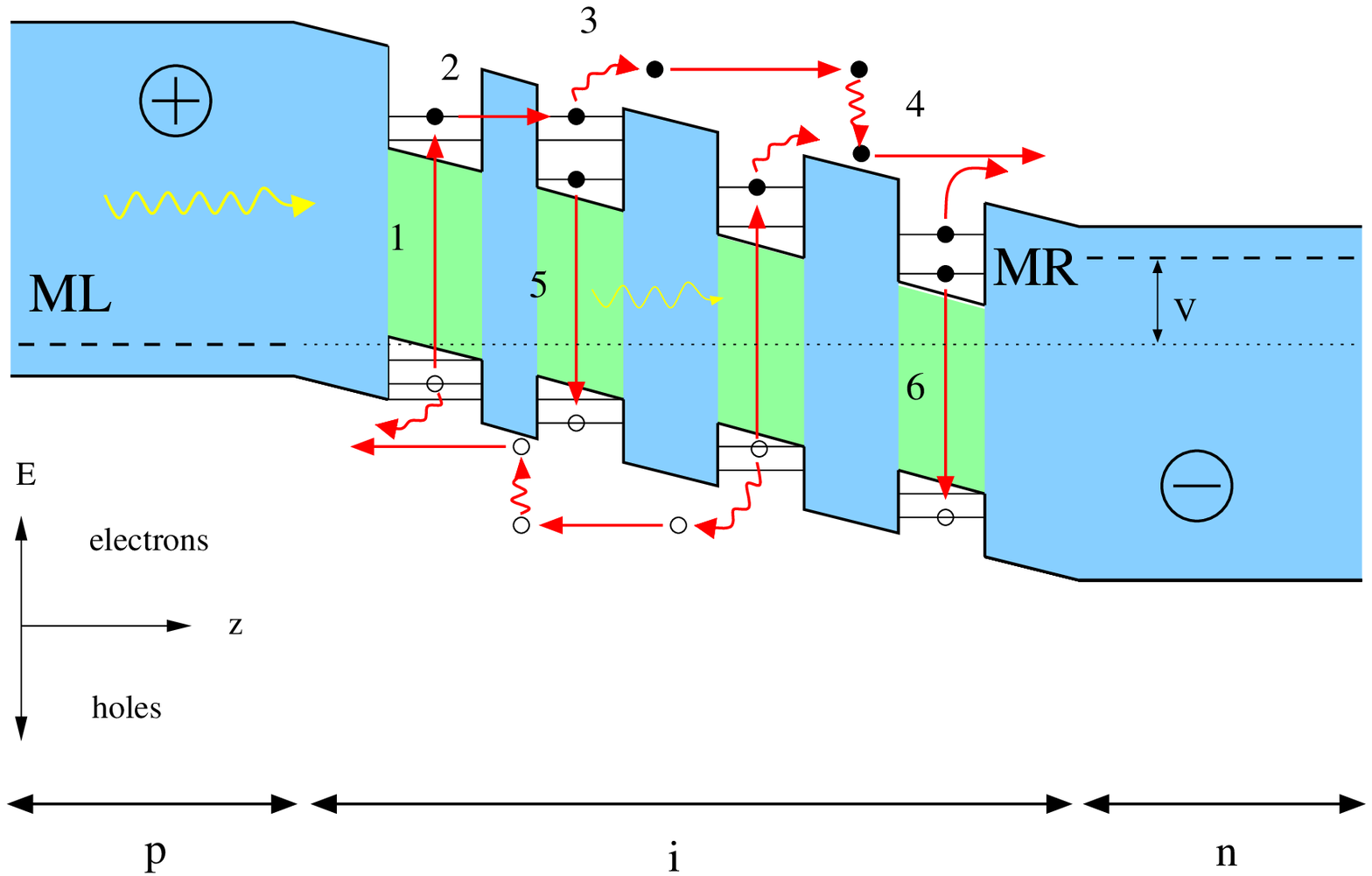}    
     \end{minipage}
\newline
\newline
\newline
\begin{minipage}{7cm}
  \flushleft{\bf Generation and recombination}
  \newline
  \newline
  \begin{tabular}{l}
  1. Photogeneration of electron-hole pairs\\
  5. Radiative recombination\\
  6. Nonradiative recombination (Auger, trap)
  \end{tabular}
  \flushleft{\bf Transport}
  \newline
  \newline
   \begin{tabular}{l}
  2. Resonant and nonresonant tunneling\\
  3. Thermal escape and sweep-out by built in field\\
  4. Relaxation by inelastic scattering (optical phonons)
  \end{tabular}
  \end{minipage}
\caption{\label{fig:qwsc}(color online) Characterizing structure and processes
of a $pin$-QWSC.}
  \end{figure}

Since the pioneering work of Barnham
and co-workers \cite{barnham:90} in the
early nineties, the potential efficiency enhancement by the introduction of quantum wells in
the intrinsic region of a
\emph{pin}-diode solar cell (Fig. \ref{fig:qwsc}) has attracted considerable
interest both from
the photovoltaic community and within a broad spectrum of fundamental research
\cite{barnham:02}.
A consistent and quantitative description of the carrier generation,
recombination, relaxation and transport processes in quantum well solar cells (QWSC)
requires the combination of a microscopic model for the electronic structure with
a formalism for quantum transport in interacting systems. 
The non-equilibrium Green's function formalism (NEGF), first introduced by
Kadanoff and Baym
\cite{kadanoff:62} and by Keldysh \cite{keldysh:65}, together with a
tight-binding
or Wannier basis meets these requirements and has been sucessfully applied to
similar systems such as quantum cascade lasers\cite{wacker_sl:02,lee_prb:02}, 
infrared photodetectors \cite{henrickson:02},
carbon nanotube photodiodes\cite{stewart:04,guo:06} or resonant tunneling diodes
\cite{lake:97}. 

The paper is organized as follows: In section II, we introduce the model Hamiltonian in a planar 
orbital basis and the procedure based on the NEGF formalism to use it in the
derivation of physical quantities.
Section III presents and discusses typical results of the theory
for a generic bipolar quantum well structure. Section V summarizes the paper and
provides an outlook to future work.
\section{\label{sec:level2}Microscopic model for QWSC}
\subsection{Hamiltonian and basis} 
The QWSC system is described in terms of the model Hamiltonian
\begin{equation}
 \hat{H}=\hat{H}_{0}+\hat{H}_{i},\quad \hat{H}_{i}=\hat{H}_{ep}+\hat{H}_{e\gamma}.
\end{equation}
$\hat{H}_{0}$ provides ballistic transport: it contains the kinetic energy, the
(bulk) band structure and band offsets, and also includes the electrostatic
potential from the solution of Poisson's equation, which corresponds to the
consideration of single species carrier-carrier scattering on the Hartree level.
The interaction part $\hat{H}_{i}$ consists of the terms $\hat{H}_{ep}$ and
 $\hat{H}_{e\gamma}$ for electron-phonon and electron-photon scattering, respectively.
Any other kind of interaction, like scattering by ionized impurities, alloy
composition inhomogeneities or interface roughness, inter-carrier-scattering
beyond the Hartree level, and all non-radiative recombination
processes like Auger or trap recombination are contained in additional terms
that will not be discussed in this paper, which is focused on the radiative
limit.

In layered semiconductors, carrier Bloch states can be represented in terms of
linear combinations of planar orbitals \cite{chang:81}
\begin{align}
  |n,({\mathbf k},k_{z})\rangle&=\sum_{\alpha,L}C_{\alpha,L}({\mathbf
    k},k_{z})|{\alpha,L,\mathbf k}\rangle,\\
    |\alpha,L,{\mathbf k}\rangle&=\frac{1}{\sqrt{N}}\sum_{{\mathbf
      R}_{\parallel}^{L}}e^{i{\mathbf k}\cdot{\mathbf R}_{\parallel}^{L}}|\alpha,L,{\mathbf
      R}_{\parallel}^{L}\rangle,
      \label{eq:pobstate}
\end{align}
where $n$ is the band index, $\alpha$ denotes a set of orthogonal localized
orbitals (e.g. $s,p_{x},p_{y},p_{z},s^{*}$ in a 10 band model for zinc-blende materials\cite{vogl:83}),
$L$ indicates
the layer, which can consist of several different atomic layers, and ${\mathbf
R}_{\parallel}^{L}$ the (transverse) location within
the layer. $N$ is a normalization factor and ${\mathbf k}$, $k_{z}$ the
transverse and longitudinal wave vectors, respectively. The corresponding field operators
are
\begin{align}
  \hat{\psi}({\mathbf r})&=\sum_{{\mathbf k},L}\sum_{\alpha}\langle {\mathbf
  r}|\alpha,L,{\mathbf k}\rangle \hat{c}_{\alpha,L,{\mathbf k}}, \\
   \hat{\psi}({\mathbf r})^{\dagger}&=\sum_{{\mathbf
   k},L}\sum_{\alpha}\langle\alpha,L,{\mathbf k}|\mathbf{r}\rangle \hat{c}^{\dagger}_{\alpha,L,{\mathbf k}},
  \label{eq:fieldop}
\end{align}
where $\hat{c}_{\alpha,L,{\mathbf k}}$ ($\hat{c}^{\dagger}_{\alpha,L,{\mathbf k}}$) is the
annihilation (creation) operator for a fermion in state $|\alpha,L,{\mathbf
k}\rangle$. 

In a planar orbital basis (POB), the
Hamiltonian for ballistic transport is expressed as
\begin{align}
\hat{H}_{0}&=\sum_{\mathbf{k}}\sum_{\alpha,\alpha'}\sum_{L,L'}\big[
t_{\alpha,L;\alpha',L'}(\mathbf{k})(1-\delta_{L',L})\hat{c}_{\alpha,L,\mathbf{k}}^{\dagger}\hat{c}_{\alpha',L',\mathbf{k}}\nonumber\\
&+D_{\alpha,L;\alpha',L}(\mathbf{k})\delta_{L,L'}\hat{c}_{\alpha,L,\mathbf{k}}^{\dagger}\hat{c}_{\alpha',L,\mathbf{k}}\big],
\end{align}
where $D$ contains the on-site energy, the intra-layer
couplings (overlap integrals) and the Hartree potential, while $t$ denotes the
inter-layer coupling.

The operator for carrier-photon interaction reads
\begin{equation}
  \hat{V}_{e\gamma}=\frac{e}{m_{0}}\hat{{\mathbf A}}\cdot\hat{{\mathbf p}},
  \end{equation}
with the quantized photon field given by 
\begin{equation}
  \hat{\mathbf{A}}({\mathbf r},t)=\frac{1}{\sqrt{V}}\sum_{\lambda{\mathbf q}}
  \sqrt{\frac{\hbar}{2\epsilon_{0}
\omega_{{\mathbf q}}}}{\mathbf e}_{\lambda{\mathbf q}}e^{i{\mathbf q}{\mathbf r}}
\left[\hat{b}_{\lambda,{\mathbf q}}(t)+
\hat{b}_{\lambda,-{\mathbf q}}^{\dagger}(t)\right],
\end{equation}
where ${\mathbf e}_{\lambda{\mathbf q}} $ is the polarization of the photon in
mode $\lambda$ and with momentum
${\mathbf q}$ created by the boson creation and annihilation operators
$\{\hat{b}^{\dagger},\hat{b}\}$, and $V$ is the absorbing volume.

In a first approach, we restrict the discussion to single-mode monochromatic
photons of energy $\hbar\omega_{\gamma}$ and use the standard dipole
approximation, which yields\cite{henrickson:02} 
\begin{align}
  \hat{{\mathbf A}}&=\mathcal{N}{\mathbf a }(\hat{b}e^{-i\omega_{\gamma} t}+\hat{b}^{\dagger}
  e^{i\omega_{\gamma} t}),\\
  \mathcal{N}&=\sqrt{\frac{\hbar\sqrt{\mu\epsilon}
  \phi_{\omega_{\gamma}}}{2N_{\gamma}\omega_{\gamma}\epsilon_{0}}},~\phi_{\omega_{\gamma}}=\frac{N_{\gamma}
  c}{V\sqrt{\mu\epsilon}}=\frac{I_{\gamma}}{\hbar\omega_{\gamma}},
\end{align}
where ${\mathbf a}$ is the polarization and $\phi_{\omega_{\gamma}}$
represents the incoming photon flux, which depends
on the intensity $I_{\gamma}$ and the photon energy, and
provides $N_{\gamma}$ photons per absorbing volume $V$ and for given 
optical properties ($\epsilon$: dielectric constant, $\mu$: magnetic
permeability). 
In the POB for a layered system, the Hamiltonian for electron-photon interaction
takes the form
\begin{align}
  \hat{H}_{e\gamma}&=\int d^{3}r\hat{\psi}^{\dagger}({\mathbf r})\hat{V}_{e\gamma}\hat{\psi}({\mathbf
 r})\\
&=\sum_{L,L'}\sum_{\alpha,\alpha'}\sum_{{\mathbf k}}M^{\gamma}_{\alpha,L;\alpha',L'}({\mathbf k})
\hat{c}^{\dagger}_{\alpha,L,{\mathbf k}}\hat{c}_{\alpha',L',{\mathbf k}}\nonumber\\
&\times(\hat{b}e^{-i\omega_{\gamma} t}+\hat{b}^{\dagger}e^{i\omega_{\gamma} t}).
\end{align}
In the dipole approximation, $\hat{\mathbf{A}}$ has no spatial dependence and thus 
\begin{equation}
M^{\gamma}_{\alpha,L;\alpha',L'}({\mathbf k})=\frac{e}{m_{0}}{\mathbf
A_{0}}\langle\alpha,L,{\mathbf k}|\hat{{\mathbf p}}|\alpha',L',{\mathbf k}\rangle,
\end{equation}
where $\mathbf A_{0}=\sqrt{\frac{\hbar}{2\epsilon_{0}V
\omega_{{\mathbf q}}}}{\mathbf a}$, $e$ is the electron charge and $m_{0}$ its
bare mass. The band structure model dependent dipole-matrix
elements for the (direct) interband transitions can be written in terms of the
tight-binding Hamiltonian as
\cite{voon:93,graf:95,boykin:01} 
\begin{align}
\langle\alpha,L,{\mathbf k}|\hat{{\mathbf p} }|\alpha',L',{\mathbf
k}\rangle=&\frac{1}{\sqrt{N}}\sum_{\mbf{R}^{L}_{\parallel},\mbf{R}^{L'}_{\parallel}}
e^{i\mbf{k}_{\parallel}\cdot(\mbf{R}^{L'}_{\parallel}-\mbf{R}^{L}_{\parallel})}\nonumber\\
&\times\langle\alpha,L,\mbf{R}^{L}_{\parallel}|\hat{{\mathbf p}
}|\alpha',L',\mbf{R}^{L'}_{\parallel}\rangle,\\ 
\langle\alpha,L,\mbf{R}^{L}_{\parallel}|{\hat{\mathbf p}
}|\alpha',L',\mbf{R}^{L'}_{\parallel}\rangle=&\frac{m_{0}}{i\hbar}\langle\alpha,L,\mbf{R}^{L}_{\parallel}|\left[\hat{\mbf{r}},\hat{H}_{0}\right]
|\alpha',L',\mbf{R}^{L'}_{\parallel}\rangle\nonumber\\
=&\frac{m_{0}}{i\hbar}(\mbf{R}^{L'}-\mbf{R}^{L})[H_{0}]_{\alpha,L;\alpha',L'},
\end{align}
where $\mbf{R}^{L}\equiv(\mbf{R}^{L}_{\parallel},L\Delta)$.
In the case of light incidence normal to the layer, the polarization is purely
transverse, and $M^{\gamma}$ becomes a scalar function of the transverse momentum.

For the interaction of carriers with phonons, which is on the level of a
coupling to an equilibrium heat bath, the harmonic approximation provides the
interaction term
\begin{equation}
	\hat{V}_{ep}=\frac{1}{\sqrt{V}}\sum_{{\mathbf q}}U_{{\mathbf q}}e^{i{\mathbf q}
     \cdot{\mathbf r}}(\hat{a}_{{\mathbf q}}+\hat{a}_{-{\mathbf q}}^{\dagger}), 
\end{equation}
where $U_{q}$ characterizes the coupling matrix elements.
In the case of a diatomic basis, as in zinc-blende compounds, the corresponding
POB interaction Hamiltonian is given by
\begin{align}
    \hat{H}_{ep}&=\int d^{3}r\hat{\psi}^{\dagger}({\mathbf 
    r})\hat{V}_{ep}\hat{\psi}({\mathbf r})\\
 &=\sum_{L,{\mathbf k}}\sum_{{\mathbf q}}\sum_{\alpha}M^{ep}_{\alpha,L}(\mathbf{q})\hat{c}_{\alpha,L,{\mathbf k}}^{\dagger}
\hat{c}_{\alpha,L,{\mathbf k-q_{t}}}\nonumber\\
 &\times(\hat{a}_{{\mathbf q}}+\hat{a}_{-{\mathbf q}}^{\dagger}).
\end{align}
where the exact form of the coupling element $M^{ep}$ depends again on the band
structure model.

\subsection{Green's functions and self-energies}
Within the planar orbital basis, the real time non-equilibrium Green's functions
are defined as the non-equilibrium ensemble averages 
\begin{align}
  G^{<}_{\alpha,L;\alpha',L'}({\mathbf
    k};t,t')&\equiv\frac{i}{\hbar}\langle \hat{c}^{\dagger}_{\alpha',L',{\mathbf
      k}}(t')\hat{c}_{\alpha,L,{\mathbf k}}(t)\rangle,\\
  G^{>}_{\alpha,L;\alpha',L'}({\mathbf
    k};t,t')&\equiv-\frac{i}{\hbar}\langle \hat{c}_{\alpha,L,{\mathbf k}}(t)\hat{c}^{\dagger}_{\alpha',L',{\mathbf
      k}}(t')\rangle,\\
  G^{R}_{\alpha,L;\alpha',L'}({\mathbf k};t,t')&\equiv\Theta(t-t')[G^{>}_{\alpha,L;\alpha',L'}({\mathbf k};t,t')\nonumber\\
    &-G^{<}_{\alpha,L;\alpha',L'}({\mathbf k};t,t')],\\
  G^{A}_{\alpha,L;\alpha',L'}({\mathbf k};t,t')&\equiv\Theta(t'-t)[G^{<}_{\alpha,L;\alpha',L'}({\mathbf k};t,t')\nonumber\\
    &-G^{>}_{\alpha,L;\alpha',L'}({\mathbf k};t,t')].
\end{align}
In steady state, the above Green's functions depend only on the time
difference $\tau=t-t'$, and it is thus possible to work with the
Fourier transform 
\begin{equation}
 G_{\alpha,L;\alpha',L'}({\mathbf
    k};E)=\int d\tau e^{iE\tau/\hbar}G_{\alpha,L;\alpha',L'}({\mathbf
    k};\tau),\quad\tau\equiv t-t'.
\end{equation}    

The effects of carrier injection and absorption by extended, highly doped
contacts acting as reservoirs are absorbed into respective
boundary self energies $\Sigma^{B}$, reflecting the openness of the system and
leading to an effective Hamiltonian of the truncated system
\cite{caroli:71,datta:95,sanvito:99}. Since the contacts form
equilibrated flat band regions, their propagating and evanescent bulk
Bloch states can be determined exactly. The boundary self energy then represents
the matching of the planar orbital states in the device to the extended lead
modes at the interface of the contacts, corresponding to a quantum transmitting
boundary method\cite{lent:90}. For instance at the left boundary ($L=1$), the
retarded boundary self energy is given by (see appendix for a detailed derivation)
\begin{align}
\Sigma_{1;1}^{RB}(\mathbf{k},E)=&t_{1;0}(\mathbf{k})\nonumber\\
&\times\big(U_{-}(\mathbf{k},E)[\Lambda_{z}(\mathbf{k},E)]^{-1}[U_{-}(\mathbf{k},E)]^{-1}\big)^{-1},
\end{align}
where $U_{-}$ specifies the transformation from localized basis to left-travelling
Bloch states and $\Lambda_{z}$ is the interlayer propagator for the
corresponding bulk modes\cite{ando:91,bouwen:95,ogawa:99}. The lesser and
greater self energies are then obtained from the broadening function
$\Gamma_{1}^{B}$ 
and the  Fermi distribution $f_{\mu_{L}}$ of the contact characterized by the
chemical potential $\mu_{L}$,
\begin{align}
\Sigma^{<B}_{1;1}(\mathbf{k},E)=&if_{\mu_{L}}(E)\Gamma^{B}_{1}(\mathbf{k},E),\\
\Sigma^{>B}_{1;1}(\mathbf{k},E)=&-i[1-f_{\mu_{L}}(E)]\Gamma^{B}_{1}(\mathbf{k},E),\\
\quad \Gamma^{B}_{1}(\mathbf{k},E)=&i[\Sigma_{1,1}^{RB}-(\Sigma^{RB})_{1,1}^{\dagger}].
\end{align}
Analogous expressions are found for the right contact.

While the boundary self energies result from an exact
treatment, interactions such as carrier-phonon and carrier-photon scattering are
included perturbatively in terms of interaction self energies $\Sigma$ on the
level of a self-consistent Born approximation (SCBA). 
The self energies for both carrier-photon and carrier-phonon are obtained from
the Fock term in second order perturbation theory for general
carrier-boson interaction. The corresponding Hartree term is neglected at the
present stage (see e.g. \onlinecite{hyldgaard:94} for an extensive discussion). In the
case of the light-matter interaction the Bose-Einstein distribution $N_{{\mathbf
q}}(\hbar\omega_{{\mathbf q}})$ in the equilibrium bosonic propagator is
replaced by the number of photons $N_{\gamma}$ present in a layer.
The lesser and greater self energies read (in full matrix notation)
\begin{align}
\Sigma_{e\gamma}^{\lessgtr}({\mathbf k};E)&=i\hbar M^{\gamma}({\mathbf k})
  \Big[N_{\gamma}G^{\lessgtr}({\mathbf k};E\mp\hbar\omega_{\gamma}\nonumber\\
  &+(N_{\gamma}+1)G^{\lessgtr}({\mathbf k};E\pm\hbar\omega_{\gamma})\Big]M^{\gamma}({\mathbf k}),\label{eq:firstse}
\end{align}
and the retarded self energy is given by
\begin{align}
 \Sigma_{e\gamma}^{R}({\mathbf k};E)&=i\hbar M^{\gamma}({\mathbf k})\Bigg[(N_{\gamma}+1)
  G^{R}({\mathbf k};E-\hbar\omega_{\gamma})\nonumber\\
  &+N_{\gamma}G^{R}({\mathbf k};
  E+\hbar\omega_{\gamma})+\frac{1}{2}[G^{<}({\mathbf k};E-\hbar\omega_{\gamma})\nonumber\\
  &-G^{<}
   ({\mathbf k};E+\hbar\omega_{\gamma})]\nonumber\\
    &+i\mathcal{P}\Bigg\{\int\frac{\d
         E'}{2\pi}\Bigg(\frac{G^{<}({\mathbf k};E-E')}{E'-\hbar\omega_{\gamma}}\nonumber\\
         &-\frac{G^{<}({\mathbf k};E-E')}{E'+\hbar\omega_{\gamma}}\Bigg)\Bigg\}\Bigg]M^{\gamma}({\mathbf k})
\end{align}
The principal value $\mathcal{P}$ in the expression for the retarded self energy
is often neglected, since it will only
contribute an energy renormalization, but not to relaxation or phase breaking.
We will adopt this approximation in the present work.

For the interactions with polar optical phonons, the self energies are then
given by (again neglecting the principal value integration in the retarded case)
\begin{align}
  \Sigma_{\alpha,L;\alpha',L'}^{\lessgtr (pop)}({\mathbf k};E)=&\sum_{{\mathbf
      q}_{\parallel}}M^{pop}(\mathbf{k},\mathbf{q}_{\parallel};L,\alpha;L',\alpha')\nonumber\\
    &\times\big[N_{LO}G^{\lessgtr}_{\alpha,L;\alpha',L'}(\mathbf{q}_{\parallel};E\mp\hbar\omega_{LO})\nonumber\\
    &+(N_{LO}+1)G_{\alpha,L;\alpha',L'}^{\lessgtr}(\mathbf{q}_{\parallel};E\pm\hbar\omega_{LO})\big]\\
     \Sigma_{\alpha,L;\alpha',L'}^{R (pop)}({\mathbf k};E)=&\sum_{{\mathbf
      q}_{\parallel}}M^{pop}(\mathbf{k},\mathbf{q}_{\parallel};L,\alpha;L',\alpha')\nonumber\\
    &\times\Big\{\big[N_{LO}G^{R}_{\alpha,L;\alpha',L'}(\mathbf{q}_{\parallel};E+\hbar\omega_{LO})\nonumber\\
    &+(N_{LO}+1)G_{\alpha,L;\alpha',L'}^{R}(\mathbf{q}_{\parallel};E-\hbar\omega_{LO})\big]\nonumber\\
    &+\frac{1}{2}[G_{\alpha,L;\alpha',L'}^{<}(\mathbf{q}_{\parallel};E-\hbar\omega_{LO})\nonumber\\
    &-G_{\alpha,L;\alpha',L'}^{<}(\mathbf{q}_{\parallel};E+\hbar\omega_{LO})]\Big\}
\label{}
\end{align}
where $N_{LO}$ is the Bose-Einstein distribution for equilibrium bosons with
energy
$E_{phon}=\hbar\omega_{LO}$ and at lattice temperature $T_{0}$. $M^{pop}$ is
a basis dependent function of the coupling parameters, spatial structure and
momentum transfer. 

For low energy
(elastic) scattering with acoustic phonons and high lattice temperature, the
expression for the equilibrium
phonon propagator can be simplified 
to provide the
(block-)diagonal, momentum independent self energies
\begin{align}
 \Sigma_{\alpha,L;\alpha',L'}^{\lessgtr,R
 (ac)}(E)=&\delta_{L,L'}M^{ac}_{\alpha;\alpha'}\sum_{\mathbf{k}}G^{\lessgtr,R}_{\alpha,L;\alpha',L'}(\mathbf{k};E).\label{eq:lastse}
\end{align}
A detailed
derivation of the electron-phonon self energies for zinc-blende structures can be
found e.g. in (\onlinecite{lake:06}).

\subsection{Quantum kinetic equations}
 Within the NEGF formalism, the steady state equations of motion for the Green's
 functions are given (in matrix notation) by the Dyson's equations
\begin{align}
   G^{R}(\mathbf{k},E)&=\left[\big(G_{0}^{R}(\mathbf{k},E)\big)^{-1}-\Sigma^{R}(\mathbf{k},E)-\Sigma^{RB}(\mathbf{k},E)\right]^{-1},\label{eq:retgf}\\
    G_{0}^{R}(\mathbf{k},E)&=\left[(E+i\eta)\mathbbm{1}-H_{0}(\mathbf{k})\right]^{-1},\\
   G^{\lessgtr}(\mathbf{k},E)&=G^{R}(\mathbf{k},E)\left(\Sigma^{\lessgtr}(\mathbf{k},E)+\Sigma^{\lessgtr B}(\mathbf{k},E)\right)G^{A}(\mathbf{k},E),\label{eq:corrf}\\
   G^{A}(\mathbf{k},E)&=(G^{R}(\mathbf{k},E))^{\dagger}.
\label{eq:GR_steadystate}
\end{align}
Together with the expressions for the self energies from boundaries and
interactions, 
and the macroscopic Poisson equation 
\begin{equation}
\epsilon_{0}\frac{d}{dz}\left[\epsilon(z)\frac{d}{dz}U(z)\right]=n(z)-p(z)-N_{dop}(z),
\label{eq:poisseq}
\end{equation}
relating the Hartree potential $U(z)$ to doping density $N_{dop}(z)$ and
the carrier densities derived
from the Green's functions, these form a closed set of
equations for the latter that have to be solved self-consistently.
To lower the computational costs, the recursive Green's function method
\cite{mackinnon:85,anda:91,ando:91} is applied, and only the first (block)off-diagonal of the
self energies is considered. 


\subsection{Carrier and current density}
The local density of states (LDOS) at layer $L$ is given by
\begin{align}
\rho_{L}(E)&=\sum_{{\mathbf k}}tr\{A_{L;L}({\mathbf k};E)\},\\
 A&=i(G^{R}-G^{A}), 
\end{align}
where $A$ is the spectral function and the trace is over orbital indices. The
averaged electron (hole) density at layer $L$ is
\begin{equation}
  n(p)_{L}=-\frac{2i}{\mathcal{A}\Delta}\sum_{{\mathbf
 k}}\int\frac{dE}{2\pi}tr\{G^{<(>)}_{L;L}({\mathbf k};E)\}.
\end{equation}
where $\mathcal{A}$ denotes the cross section area and $\Delta$ the layer thickness.
The current density passing from layers
$L$ to $L+1$ is 
\begin{align}
 J_{L}^{n(p)}&=\frac{2e}{\hbar \mathcal{A}}\sum_{{\mathbf k}}\int\frac{dE}{2\pi}tr\{t_{L;L+1}G^{<(>)}_{L+1;L}({\mathbf k};E)\nonumber\\
 &-t_{L+1;L}G^{<(>)}_{L;L+1}({\mathbf k};E)\}\label{eq:current}.
\end{align}

\subsection{Absorption} 
The absorption of a given layer in a illuminated heterostructure can be derived
in terms of the microscopic interband polarization $\Pi_{cv}$\cite{faleev:02},
\begin{equation}
\alpha_{L}(\hbar\omega)=-\frac{4\pi}{\sqrt{\epsilon}\mathcal{A}\Delta 
c\hbar\omega}\Im\{(\Pi_{cv}^{R})_{L;L}(q=0,\hbar\omega)\},
\end{equation}
where 
\begin{align}
\Im\{\Pi_{cv}^{R}(\mathbf{k},E)\}&=-\frac{i}{2}\left[\Pi^{>}_{cv}(\mathbf{k},E)-\Pi^{<}_{cv}(\mathbf{k},E)\right]\\
&=-\frac{i}{2}\left[\Pi^{<}_{vc}(\mathbf{k},E)-\Pi^{<}_{cv}(\mathbf{k},E)\right].
\end{align}
and 
\begin{align}
\Pi_{cv}^{\lessgtr}(0,E)&=-2i\int\frac{dE'}{2\pi\hbar}\frac{dq}{2\pi}q|M^{\gamma}_{cv}(q)|^{2}\nonumber\\
&\times G^{\lessgtr}_{c}(\mathbf{q},E')G^{\gtrless}_{v}(\mathbf{q},E'-E).
\end{align}
The incoming photon flux $\phi_{\gamma}$, after passing through layers $L_{1},L_{2},..,L_{N}$, is
reduced by the absorptivity 
\begin{align}
a_{\gamma}=1-\exp[-\sum_{n=1}^{N}\alpha_{L_{n}}(\hbar\omega_{\gamma})\Delta],
\quad\phi_{\gamma,abs}=\phi_{\gamma}a_{\gamma}
\end{align}
where $\phi_{\gamma,abs}$ is the absorbed photon flux.
\subsection{Computational scheme}
\begin{figure}[h!]
 \begin{center}
   \psfrag{Poiss}[][][0.9]{Poisson's equation}
   \psfrag{U}[][][0.9]{ $U$}
   \psfrag{output}[][][0.9]{Output}
   \psfrag{GR}[][][0.9]{$G^{R}$}
   \psfrag{corrfn}[][][0.9]{$G^{\lessgtr}$}
   \psfrag{SigB}[][][0.9]{$\Sigma_{L,R}^{R,\lessgtr B}$}
   \psfrag{Scatt}[][][0.9]{$\Sigma^{\lessgtr},~\Sigma^{R}$}
   \psfrag{conv}[][][0.9]{converged?}
   \psfrag{dens}[][][0.9]{$n,~p,~J$}
   \psfrag{yes}[][][0.9]{yes}
   \psfrag{no}[][][0.9]{no}
   \psfrag{Dyson}[][][0.9]{Dyson's equation}
   \psfrag{self}[][][0.9]{self-energy equations}
   \psfrag{eqn}[][][0.9]{equation}
   \psfrag{kel}[][][0.9]{Keldysh equation}
     \includegraphics[width=8cm]{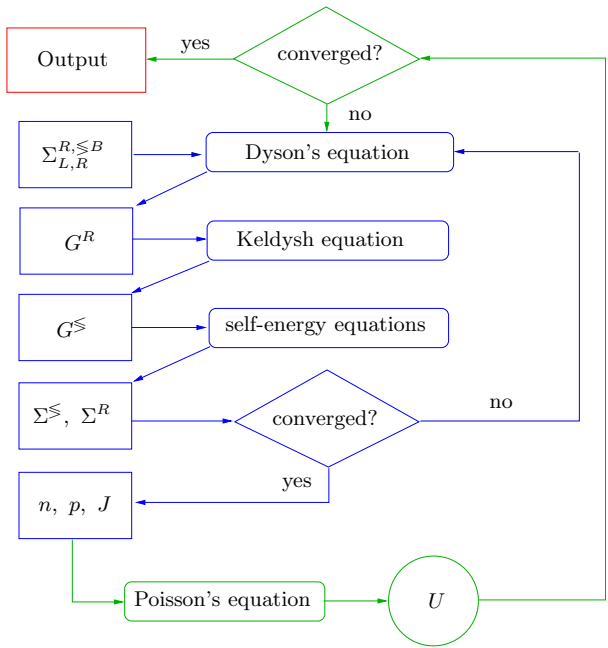}
     \caption{\label{fig:compscheme} Computational scheme for the calculation of
     physical quantities from Green's functions and self energies. The inner
     self-consistency loop connects the equations for the Green's functions and
     the self energies, while the outer loop provides the update of the Hartree
     potential from the solution of Poisson's equation.}
     \end{center}
  \end{figure}

After choosing an initial potential profile (e.g. from the depletion
approximation), the boundary self energies are calculated and used in the Dyson
equation \eqref{eq:retgf} for the retarded Green's function $G^{R}$, followed by the
evaluation of the
Keldysh equation \eqref{eq:corrf} for the correlation functions $G^{\lessgtr}$. These Green's
functions provide
an update of the scattering self energies $\Sigma^{R,\lessgtr}$ 
\eqref{eq:firstse}-\eqref{eq:lastse} and the values of density and
current. The new self energies are again used in the equations for the Green's
functions, and this self-consistency iteration is continued until convergence is reached. 
Since the calculation of (photo)current is central to this work, its convergence
is used as the aborting condition instead of that of the Green's functions or
self energies\footnote{From Eq. \eqref{eq:current} follows that convergence of the current
depends on the convergence of the real part of the off-diagonal elements of the
correlation functions.}. To obtain the built-in electric field, but also in
cases where charging 
effects cannot be neglected (e.g. deep wells at large bias), 
Poisson's equation is solved in an additional self-consistency loop using the
densities from the NEGF and providing an update to the Hartree potential in the ballistic
Hamiltonian. The computational
scheme is represented in Fig. \ref{fig:compscheme}. 

\section{Results and Discussion}
The following results for generic single quantum well (SQW) pin-diodes were obtained
using the
two band $sp_{z}$-Hamiltonian with parabolic and isotropic tranverse dispersion
discussed in the appendix.
Table \ref{tab:table1} shows the set of microscopic parameters used in the
simulations. To lower the computational burden, short structures of 70-100 nm
with reduced energy gaps of 0.5 eV (well) and 0.9 eV (barrier) are
investigated. The band offsets of barrier and well material are chosen
to resemble those of the GaAs-Al$_{x}$Ga$_{1-x}$As System with $x\sim 0.3$, i.e.
0.25 eV for the
conduction band offset and 0.15 eV for valence band discontinuity. The
contacts are made of 50 monolayers (ML) of high bandgap material with
strong doping ($N_{d,a}=10^{18}$ cm$^{-3}$). Between contact and active device,
intrinsic buffer regions of 60 ML are inserted. The calculations are performed
at 300 K, the illumination intensity is 1000 W/m$^{2}$ ($\sim$ 1 sun) and the
cross section is
$\mathcal{A}=1$ cm$^{2}$. The photon energies are chosen in the range of the
confinement level separation between the two band gap values, such that the
contact and lead regions are non-absorbing.

\begin{table}[b]
\caption{\label{tab:table1} Material parameters used in simulations}
\begin{ruledtabular}
\begin{tabular}{lccl}
&barrier &well &\\
\hline
 $E_{s}$& 0.75 & 0.5 & $s$-orbital onsite energy\\
 $E_{p}$& -0.15 & 0&  $p_{z}$-orbital onsite energy\\
 $V_{sp}$&2.8& 2.5& layer coupling element\\
$m^{*}_{Cb}/m_{0}$& 0.1087 & 0.067&effective mass in conduction band\\
$m^{*}_{Vb}/m_{0}$& 0.29 & 0.23&effective mass in valence band\\
$\epsilon$&12.2&13.1&dielectric constant\\
$\mu$&1&1&magnetic permeability
\end{tabular}
\end{ruledtabular}
\end{table}

\subsection{Local density of states}

\begin{figure*}[htbp]
\begin{center}
\includegraphics[width=19cm]{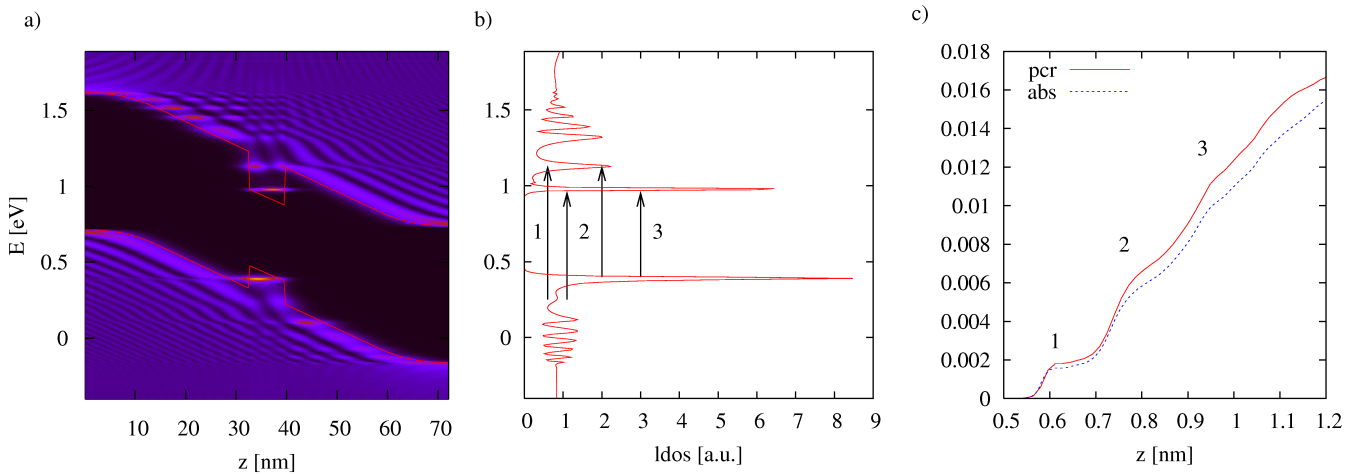}
\caption{\label{fig:ldos25ML}(color online) a) Local density of states (LDOS) at
$\mathbf{k}=0$ for a 25 ML SQW pin-diode at
 V$_{bias}=-0.01$ V: quantum confinement leads to the formation of quasi-bound
 states and higher transmission resonances in the well region, in addition to
 the stripe-like interference pattern due to the built-in field (strong band
 bending); b) LDOS at the well center and optical transitions between
 confinement levels: the quasi-bound states near the well edge show the
 characteristic broadening
 associated with shorter carrier dwell time, as compared to the sharp deep and 
 strongly-bound states; c) Photocurrent response
 (pcr) and absorptivity (abs): step like and square root like dependence on
 the photon energy below and above the higher band gap value, reflecting the
 density of the states participating in the corresponding optical
 transitions, i.e. confinement level to confinement level, confinement level
 to quasi-continuum and quasi-continuum transitions, respectively.}
\end{center}
\end{figure*}

Since the system is open, there are no true
bound states, and the formalism considers only states contributing to current,
i.e. connected to extended states with finite amplitude in the contacts. Fig.
\ref{fig:ldos25ML}a shows the local density of states (LDOS) for a 25 ML well at
$\mathbf{k}=0$. In this case,
two sharp confinement levels are present and contribute to the photocurrent. The
high lying state is only weakly bound and broadened, corresponding to a
faster carrier escape as compared to the more strongly bound and
sharper low lying state.
In the case of strong
scattering, phonon satellite peaks form next to the confinement level peaks, 
as visible in the cut of the LDOS through the center of the well (Fig. 
\ref{fig:ldos25ML}b).
In addition to the confined states, there is a variety of quasi-bound states and
transmission resonances above
the well, which influence the photovoltaic properties of the
structure and might explain the enhanced absorption of QWSC observed at photon
energies above the higher bandgap. One can further observe a kind of ``notch''
states between
well and the corresponding contacts, as are usually observed in the presence of
barriers. If scattering in the leads is neglected, a stripe type interference 
pattern forms due to reflection of carriers injected below the band edge at the
contacts, above which the LDOS acquires the expected uniform value of the
quasi-continuum, which however is still affected by the presence of the well. 

\subsection{Optical transitions, absorption and photocurrent response}
The different optical transitions between confined states, quasi-bound states,
higher resonances and the continuum can be identified in the photocurrent
response (Fig. \ref{fig:ldos25ML}c), which at short circuit conditions
corresponds 
to the external quantum efficiency, i.e. the short circuit current normalized by
the incoming photon
flux. Since the devices considered in this investigation are short,
photocurrent
is limited by the absorption, and it is therefore essential to normalize physical
quantities to the absorptivity (Fig. \ref{fig:ldos25ML}c) in order to allow a
comparison of different structures. 

\subsection{Current spectrum and IV-characteristics}
\begin{figure*}
\begin{center}
\includegraphics[width=19cm]{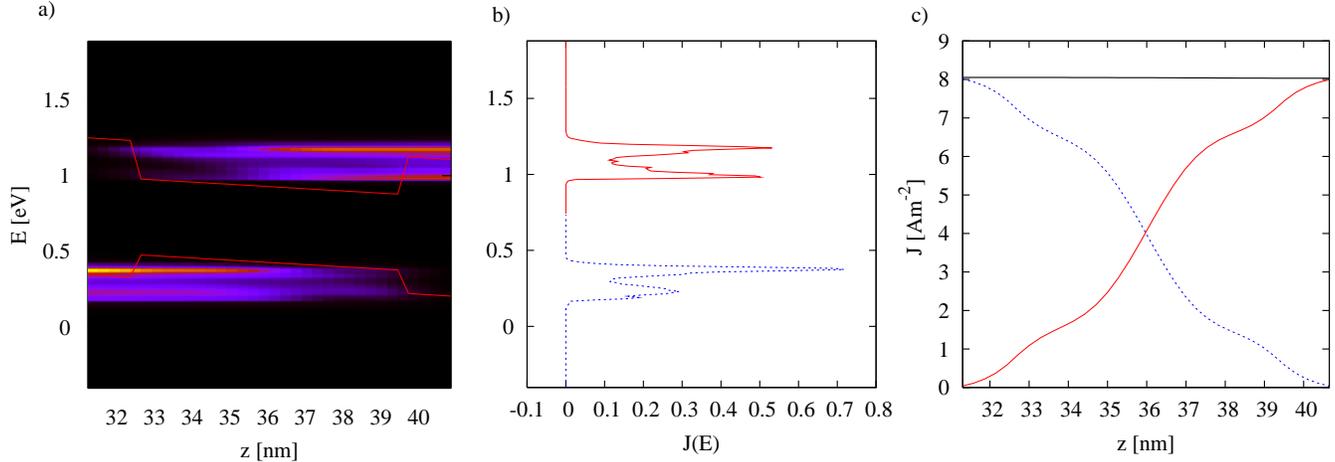}
\caption{\label{fig:lc}(color online) a) Spatially resolved photocurrent
spectrum (at zero bias voltage) in the QW region and b) at the
interface to the $n$-contact (electrons) and to the $p$-contact (holes):  the
spectrum reflects the joint density of states for the contributing transitions
between the confinement levels, modified by the probability for escape,
which is suppressed in the case of the deep electronic level. c) Electron and
hole components of the photocurrent grow towards the respective contacts, while
the total current is conserved. In the dark, the bands are uncoupled, and 
current is conserved for the two carrier species separately.}
\end{center}
\end{figure*}

There are two contributions to the total current in illuminated QW pin-diodes:
dark current, corresponding to the diode current driven by an applied external 
bias, and the photocurrent originating from the photogeneration of
electron-hole pairs. Resolution in space and energy of the current in the QW
region (Fig. \ref{fig:ldos25ML}c) allows the distinction between the two
components. The
diode current occupies a narrow region above the band edge at the contacts, it
is constant over the hole device and
its spectrum reflects the density of states and the distribution of the
carriers in the contact reservoirs
from which they are injected, broadened by scattering with acoustic phonons, and
relaxed towards lower energies by interaction with polar optical phonons. In the
absence of interband recombination, the
diode current is conserved for electrons and holes separately. Photocurrent, on
the other hand, is driven by the excitation of carriers from the opposite band,
and current conservation\footnote{The observed current
conservation is an intrinsic property of the self-consistent calculation of the
interaction self energies.} thus holds only for the sum of electron and hole
contributions, but not for the separate components, which increase towards the
respective contacts (Fig. \ref{fig:lc}c) and differ also in their spectrum (Fig. \ref{fig:lc}b). 
The photocurrent spectrum reflects the joint density
of states of the dominant transition between confinement levels. Unlike the LDOS in the well, 
the current spectrum shows a strong asymmetry between electrons and holes: in
the conduction band well, the main contribution to current comes from the higher
level, while it is the lower one that dominates the current in the valence band
well. This demonstrates the impact of carrier escape probability on the current,
the latter no longer being characterized by the LDOS alone as in bulk
structures.

\begin{figure}
\begin{center}
\includegraphics[width=8.5cm]{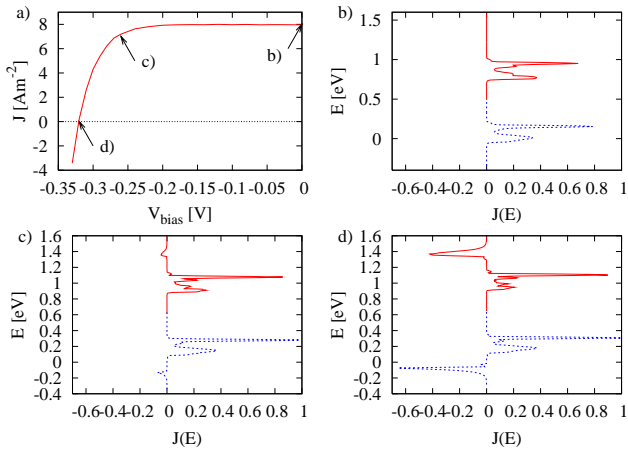}
\caption{\label{fig:iv}(color online) a) IV-characteristics for a 25 ML SQW
structure and the current spectrum at the lead-device interface for b) 0 V
(short circuit conditions), c) -0.26 V (near the maximum power point), d) -0.32
V (near the open circuit voltage). The spectrum of the exponentially increasing
diode current reflects the density of states and the distribution of the
carriers in the bulk contacts, modified
by the effects of relaxation due to inelastic scattering in the active region, which
leads to the formation of phonon satellites (weakly recognizable near the band
edge).}

\end{center}
\end{figure}

Fig. \ref{fig:iv}a shows the current-voltage characteristics for the 25 ML SQW
structure. Near short circuit conditions (Fig. \ref{fig:iv}b), current is purely
photocurrent. At increasing bias, the diode current evolves
exponentially (Fig. \ref{fig:iv}c,d), showing the
specific spectrum of the injected carriers
and the effects of scattering in
terms of phonon satellite peaks towards the band edge. The spectrum of the
photocurrent is modified due to the Stark effect. 

\section{Summary and conclusions}
We presented a microscopic model for the consistent description of generation and
transport processes in semiconductor quantum well structures under monochromatic
illumination and in the radiative limit. Based on the NEGF formalism
for a tight-binding Hamiltonian, it provides access to non-equilibrium phenomena
in quantum confined structures subject to interactions and therefore supports the
investigation of the microscopic processes governing the physics of quantum well
solar cells. 
The insights into the photovoltaic performance of specifically coupled
multi-quantum-well structures, gained from the application of the presented
approach, are the subject of current investigations and will be published
elsewhere. 
Future work will also include a microscopic treatment of the main nonradiative
recombination processes, which are Auger and trap recombination. For comparison
with experiment, a more realistic band structure model will be used. In order to
account for optical processes in extended structures, such as photon recycling,
the spatial variation of the light intensity needs to be considered, which can
be accomplished by the
solution of an additional Dyson equation for the photon propagator containing the
microscopic polarization function. Investigations of hot carrier effects will
require a corresponding treatment of the phonons in the quantum region. 

\begin{acknowledgments} 
Helpful discussions with Dr. Mathieu Luisier from IIS at ETH Zurich are
gratefully acknowledged.

\end{acknowledgments}

\appendix

\section{Two band tight-binding model}
\begin{figure}[h!]
   \psfrag{a}[][][0.8]{c}
   \psfrag{c}[][][0.8]{a}
   \psfrag{Lm1}[][][0.8]{$L-1$}
   \psfrag{L}[][][0.8]{$L$}
   \psfrag{Lp1}[][][0.8]{$L+1$}
   \psfrag{d1}[][][0.8]{$\Delta=a_{L}/4$}
   \psfrag{d2}[][][0.8]{$\Delta=a_{L}/2$}
   \psfrag{Uac}[][][0.8]{$U_{ac}$}
   \psfrag{Uca}[][][0.8]{$U_{ca}$}
   \psfrag{Vac}[][][0.8]{$V_{ac}$}
   \psfrag{Vca}[][][0.8]{$V_{ca}$}
   \psfrag{Ec}[][][0.8]{$E_{s}$}
   \psfrag{Ea}[][][0.8]{$E_{p}$}
    \begin{center}
      \includegraphics[height=4cm]{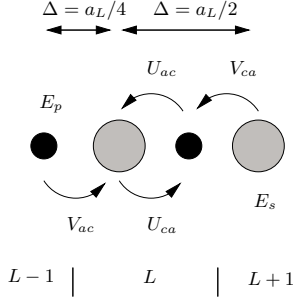}
      \caption{\label{fig:binary}Tight binding elements for a zinc-blende
      lattice in (001) direction.}
    \end{center}
    
  \end{figure}

The simplest
tight-binding model to
describe the conduction and valence band structure of III-V
semiconductors like e.g. GaAs is the diatomic model with a two-orbital
basis\cite{boyikin:96_diatom}. In this model, the
two-band dispersion is reproduced approximately by placing an $s$-type orbital
on the cation (Ga) and a
$p_{z}$-type orbital on the anion (As).
Fig. \ref{fig:binary} shows a projection of the
zinc-blende lattice onto the (001) direction, with the corresponding intra-
and interlayer couplings $U_{ac}=U_{ca}=V_{sp}$ and $V_{ac}=V_{ca}=-V_{sp}$, and
the "on-layer'' energies $E_{c}=E_{s}$ and $E_{a}=E_{p}$.
For further simplification, the transverse band structure is
approximated by an isotropic parabolic dispersion, with the electronic
effective mass
replaced by the hole effective mass in the case of the valence band.
What remains in the direction of propagation is a $sp_{z}$ model with ${\mathbf
k}=0$, which is equivalent to a linear chain of atoms, with the anion layers
at $z=na_{L}/2$ the cation layers at  $z=(na_{L}/2+a_{L}/4)$, where $n\in\mathbbm{N}$ labels
the monolayer of thickness $\Delta=a_{L}/2$, with $a_{L}$ the lattice constant of the
binary compound (i.e. 5.65 $\mathring{A}$ in the case of GaAs).
In terms of the parameters introduced above, the elements
of the TB-Hamiltonian
\begin{equation}
	H_{TB}=\begin{pmatrix}
  \ddots&\ddots&\ddots&&\\
  t_{L-1L-2}&D_{L-1}\mathbbm{1}_{b}&t_{L-1L}&&\\
  &t_{LL-1}&D_{L}\mathbbm{1}_{b}&t_{LL+1}&\\
  &&t_{L+1L}&D_{L+1}\mathbbm{1}_{b}&t_{L+1L+2}\\
  &&\ddots&\ddots&\ddots
  \end{pmatrix},
\end{equation}
including parabolic transverse energy,
are given by
\begin{align}
&D_{L}\equiv H_{nn}=\begin{pmatrix}
                   E_{s}+\frac{\hbar^{2} k^{2}}{2m^{*}_{el}}&-V_{sp}\\
                   -V_{sp}&E_{p}-\frac{\hbar^{2} k^{2}}{2m^{*}_{hl}}
                   \end{pmatrix},\\
&t_{LL+1}\equiv H_{nn+1}=\begin{pmatrix}
                   0&0\\
                   V_{sp}&0
                   \end{pmatrix},~
t_{LL-1}\equiv H_{nn-1}=\begin{pmatrix}
                   0&V_{sp}\\
                   0&0
                   \end{pmatrix}.
\end{align}
where $m^{*}_{el/hl}$ is the effective electron and hole mass, respectively.
This yields the bulk
Hamiltonian
\begin{equation}
H({\mathbf k},k_{z})=\begin{pmatrix}
                     E_{s}+\frac{\hbar^{2} k^{2}}{2m^{*}_{el}}& 2 i
                     V_{sp}\sin(k_{z}\frac{a_{L}}{4})\\
                      -2 i V_{sp}\sin(k_{z}\frac{a_{L}}{4})&E_{p}-\frac{\hbar^{2} k^{2}}{2m^{*}_{hl}}
                     \end{pmatrix}.
\end{equation}
which for $k=0$ gives rise to the dispersion relation
\begin{align}
&\det[H(k_{z})-E]=0\quad\Rightarrow E(k_{z})=\frac{1}{2}\Big[E_{p}+
E_{s}\nonumber\\
&\pm\sqrt{(E_{p}-E_{s})^{2}+16V_{sp}^{2}\sin^{2}\left(k_{z}\frac{a_{L}}{4}\right)}\Big].
\label{eq:2banddispersrel}
\end{align}
For the integration over
transverse momentum, the isotropic one dimensional approximation
\begin{equation}
\sum_{{\mathbf k}}\approx
\frac{\mathcal{A}}{(2\pi)^{2}}\int_{BZ_{\parallel}}\d^{2}k\approx\frac{\mathcal{A}}{2\pi}\int\d k
k,
\label{eq:isotropic_approx}
\end{equation}
is used, where $A$ is the device cross section, $BZ_{\parallel}$ is the projected
Brillouin
zone and $k=|{\mathbf k}|$ is the absolute value of the transverse momentum.

The tight-binding parameters are related to the longitudinal effective mass $m^{*}_{z}$
through the longitudinal dispersion relation, as
$m^{*}_{z}=\frac{\hbar^{2}}{m_{0}}\left[\ppfq{E_{z}}{k_{z}}\right]^{-1}$, with $E_{z}(k_{z})$ resulting from
the secular equation 
\begin{equation}
\det[H_{\perp}(k_{z})-E_{z})]=0,\quad
H_{\perp}(k_{z})=H({\mathbf k},k_{z})-\frac{\hbar^{2}
k^{2}}{2m^{*}_{\parallel}}.
\end{equation}
From Eq. \eqref{eq:2banddispersrel}, one
finds the relation between the effective mass at the $\Gamma$-point and the
coupling element $V_{sp}$,
\begin{equation}
m^{*\Gamma}_{z}=\frac{\hbar^{2}}{m_{0}}
\left(\frac{a^{2}V_{sp}^{2}q^{2}}{2|E_{s}-E_{p}|q}\right)^{-1}\Rightarrow
V_{sp}=\frac{\hbar}{a}\sqrt{\frac{2 E_{g}}{m^{*\Gamma}_{z}m_{0}q}},
\end{equation}
where $E_{g}=|E_{s}-E_{p}|$ is the energy gap.

\section{Boundary self energies for multiband tight-binding transport models}

To properly model the effect of semi-infinite bulk at the lead-device
interface, the interface Green's function has to be linked to the
propagating and evanescent states in the leads. The total electron
wave function expressed in terms of the Bloch sum of the anion
(a) and cation (c) states as a linear combination of planar orbitals 
$|\alpha,L,{\mathbf k}\rangle$ 
is given by Eq. \eqref{eq:pobstate}.
In the planar orbital basis, projecting onto the atomic orbitals $\alpha'$ located at layer L, 
the Schr\"odinger equation for the contact Bloch
states reads
\begin{align}
  \sum_{\alpha}\langle\alpha',L,{\mathbf
    k}|\bar{H}|\alpha,k_{z}\rangle&=0,\\
  \langle\alpha',L,{\mathbf
    k}|\bar{H}|\alpha,k_{z}\rangle&\equiv \langle\alpha',L,{\mathbf
    k}|H|\alpha,k_{z}\rangle\nonumber\\&-E\langle\alpha',L,{\mathbf
    k}|\alpha,k_{z}\rangle.  
\label{eq:localorbSG}
\end{align}
For a tight-binding Hamiltonian coupling $m$ neighboring layers, which is of the form
\begin{equation}
  \bar{H}({\mathbf k},k_{z})=\sum_{\sigma=-m}^{m}\bar{H}^{\sigma}({\mathbf k})e^{i\sigma k_{z}\Delta},
\label{}
\end{equation}
where $\bar{H}^{\sigma}({\mathbf k})$ represents a matrix which
couples a given layer to the $\sigma$-th neighboring layer and $\Delta$ is the layer spacing, and
defining
\begin{equation}
  C^{\sigma}_{\alpha}\equiv e^{i\sigma k_{z}\Delta}C_{\alpha}, \quad \sigma=-m,..,m,
\label{}
\end{equation}
Eq.\eqref{eq:localorbSG} can be written as
\begin{equation}
  \sum_{\sigma=-m}^{m-1}\bar{H}^{\sigma}C^{\sigma}+\bar{H}^{m}e^{ik_{z}\Delta}C^{m-1}=0, 
\label{}
\end{equation}
where it was used that $C^{m}=e^{ik_{z}\Delta}C^{m-1}$.\newline
For a nearest neighbor Hamiltonian ($m=1$), the projected Schr\"odinger
equation is recast into
\begin{equation}
  \bar{H}^{\sigma-1}C^{\sigma-1}+\bar{H}^{\sigma}C^{\sigma}+\bar{H}^{\sigma+1}C^{\sigma+1}=0, 
\label{}
\end{equation}
which, using $C^{\sigma\pm 1}=e^{\pm ik_{z}\Delta}C^{\sigma}$, can be
written as
\begin{equation}
  \bar{H}^{\sigma-1}e^{-ik_{z}\Delta}C^{\sigma}+\bar{H}^{\sigma}C^{\sigma}+\bar{H}^{\sigma+1}e^{ik_{z}\Delta}C^{\sigma}=0. 
\label{}
\end{equation}
This equation can then be transformed into an eigenequation for the
propagation factors $\lambda=e^{ik_{z}\Delta}$ and the lead Bloch states in
local orbital basis:
\begin{equation}
  TC_{L}=\lambda C_{L}\equiv C_{L+1}
\label{eq:transfer}
\end{equation}
with $C_{L}=\left(\begin{array}{c}
    C_{a}\\
    C_{c}
    \end{array}
    \right)$
and $T=T_{c}T_{a}$, where $T_{a}$ and $T_{c}$ are the atomic layer transfer matrices
defined as
\begin{align}
  T_{b}&=\begin{pmatrix}
\label{}-\left[H_{l,l-1}^{(b)}\right]^{-1}\left[H_{l,l}^{(b)}\right]&-\left[H_{l,l-1}^{(b)}\right
]^{-1}\left[H_{l,l+1}^{(b)}\right]\\
{\mathbbm 1}&{\mathbf 0}
\end{pmatrix},\nonumber\\&(b=a,c)
\label{}
\end{align}
with the matrix elements given by ($l$ denotes the \emph{atomic} layer)
\begin{align}
  H_{l,l-1,\alpha,\alpha'}^{(b)}&=\langle{\alpha,l,\mathbf
    k}|H|\alpha',l-1,{\mathbf k}\rangle,\\
  H_{l,l,\alpha,\alpha'}^{(b)}&=\langle {\alpha,l,\mathbf
    k}|H|\alpha',l,{\mathbf k}\rangle-E\delta_{\alpha,\alpha'},\\
  H_{l,l+1,\alpha,\alpha'}^{(b)}&=\langle{\alpha,l,\mathbf
    k}|H|\alpha',l+1,{\mathbf k}\rangle.
\label{}
\end{align}
The eigenstates $\chi$ and eigenvalues $\lambda=e^{ik_{z}\Delta}$ of
Eq. \eqref{eq:transfer} correspond to the bulk modes propagating (real
$k_{z}$) or decaying (complex $k_{z}$) to the left ($\Re(k_{z})<0$) and to
the right ($\Re(k_{z})>0$), respectively. For an $N_{b}$-band model with a
two atom basis, there are $N_{b}/2$ states $\chi_{\nu}$ propagating or decaying
to the right $(\nu=+)$ and to the left $(\nu=-)$, respectively. At a given layer $L$, 
the components for left- and right travelling waves
can be written as
\begin{equation}
  {\mathbf C}_{L\pm}=U_{\pm}{\mathbf C}_{\pm},
\label{}
\end{equation}
where ${\mathbf C}_{\nu}$ is a vector containing the expansion coefficients,
and
\begin{equation}
  U_{+}=\left(\begin{array}{ccc|ccc}
    \chi_{+}^{(a)1}&\hdots&\chi_{+}^{(a)N_{b}/2}&&{\mathbf 0}&\\
    \hline
    &{\mathbf 0}&&\chi_{+}^{(c)1}&\hdots&\chi_{+}^{(c)N_{b}/2}
\end{array}\right).
\label{}
\end{equation}
The corresponding expression for the adjacent layer $L+1$ is
\begin{equation}
  {\mathbf C}_{L+1\pm}=U_{\pm}\lambda_{z}^{\pm1}{\mathbf C}_{\pm},
\label{}
\end{equation}
with the propagation matrix
\begin{equation}
  \Lambda_{z}=
\left(\begin{array}{ccc|ccc}
    e^{ik_{z}^{1}\Delta}&&0&&&\\
    &\ddots&&&{\mathbf 0}&\\
    0&&e^{ik_{z}^{N_{b}/2}\Delta}&&&\\
      \hline
     &&&e^{ik_{z}^{1}\Delta}&&0\\
      &{\mathbf 0}&&&\ddots&\\
      &&&0&&e^{ik_{z}^{N_{b}/2}\Delta}
\end{array}\right)
\end{equation}
The relation between the two layers follows as
\begin{equation}
  {\mathbf C}_{(L+1)\pm}=F_{\pm}{\mathbf C}_{L\pm}
\label{eq:propagator}
\end{equation}
with
\begin{equation}
  F_{\pm}=U_{\pm}\Lambda_{z}^{\pm1}U_{\pm}^{-1}.
\label{}
\end{equation}
Relation \eqref{eq:propagator} can be used to derive the retarded Green's
function $g^{rR}$ at the right boundary ($L=1$) of the uncoupled semi 
infinite left lead, i.e. for the case where all the couplings to the right are
set to zero. The equation 
\begin{equation}
\left[(E+i\eta)\mathbbm{1}-H_{0}\right]g^{rR}=\mathbbm{1}
\end{equation}
yields for the boundary element
\begin{equation}
  \left[(E+i\eta){\mathbbm 1}-D_{1}\right]g^{rR}_{1;1}-t_{1;1}g^{rR}_{0;1}=0.
\label{}
\end{equation}
Eq. \eqref{eq:propagator} provides the relation
\begin{equation}
  g^{rR}_{0;1}=F_{-}^{-1}g^{rR}_{1;1},
\label{}
\end{equation}
which determines the left lead boundary Green's function in terms of the bulk
modes as
\begin{align}
  g^{rR}_{1;1}&=[E{\mathbbm 1}-D_{1}-t_{1;0}F_{-}^{-1}]^{-1}\\
  &\equiv [E{\mathbbm 1}-D_{1}-\Sigma_{1;1}^{RB}]^{-1},
\end{align}
and providing thus an expression for the (left) retarded boundary self energy $\Sigma_{1;1}^{RB}$.

\newpage

\bibliographystyle{aiptitle}
\bibliography{negf,qwsc,bandstructure,generation,qtbm}

\end{document}